\documentclass[aps, letterpaper, 11pt, onecolumn,
  floatfix, fleqn, notitlepage, endfloats,
  amsmath, amssymb, amsfonts,
  preprintnumbers, showpacs, showkeys]{revtex4-1}

\usepackage[colorlinks=true, linkcolor=blue, citecolor=blue,
            filecolor=blue, urlcolor=blue]{hyperref}
\usepackage{dcolumn, graphicx, moreverb}
\newcolumntype{.}{D{.}{.}{-1}} 
\usepackage[american]{babel}

\usepackage{mathptmx}
\usepackage[T1]{fontenc}

\newcommand{\be}{\begin{equation}}
\newcommand{\ee}{\end{equation}}
\newcommand{\bea}{\begin{eqnarray}}
\newcommand{\eea}{\end{eqnarray}}

\newcommand{\q}[2]{\ensuremath{#1\ \mathrm{#2}}} 
\newcommand{\code}[1]{\textsc{#1}} 
\newcommand{\vol}[1]{\textbf{#1}} 
\newcommand{\ve}[1]{\ensuremath{\mathbf{#1}}} 
\newcommand{\ohm}{\ensuremath{\Omega}} 

\newcommand{\Epkin}{\ensuremath{T_p}} 
\newcommand{\pemit}{\ensuremath{\epsilon_p}} 
\newcommand{\sigmap}{\ensuremath{\sigma_p}} 
\newcommand{\betastar}{\ensuremath{\beta^*}} 
\newcommand{\Bg}{\ensuremath{B_g}} 
\newcommand{\Bm}{\ensuremath{B_m}} 
\newcommand{\Bc}{\ensuremath{B_c}} 
\newcommand{\kompr}{\ensuremath{k}} 
\newcommand{\Vca}{\ensuremath{V_{ca}}} 
\newcommand{\modrise}{\ensuremath{\tau_\mathrm{mod}}} 
\newcommand{\modrep}{\ensuremath{f_\mathrm{mod}}} 
\newcommand{\rgi}{\ensuremath{r_{gi}}} 
\newcommand{\rgo}{\ensuremath{r_{go}}} 
\newcommand{\rmi}{\ensuremath{r_{mi}}} 
\newcommand{\rmo}{\ensuremath{r_{mo}}} 
\newcommand{\perv}{\ensuremath{P}} 
\newcommand{\Ie}{\ensuremath{I_e}} 
\newcommand{\Lint}{\ensuremath{L}} 

\begin{document}

\title{Conceptual design of hollow electron lenses for beam halo
  control \\ in the Large Hadron Collider}

\thanks{Fermilab is operated by Fermi Research Alliance, LLC under
  Contract No.~DE-AC02-07CH11359 with the United States Department of
  Energy.  This work was partially supported by the US DOE LHC
  Accelerator Research Program (LARP) and by the European FP7 HiLumi
  LHC Design Study, Grant Agreement 284404.}

\author{G.~Stancari} \email[Email:]{$\langle$stancari@fnal.gov$\rangle$.}

\author{V.~Previtali}
\author{A.~Valishev}
\affiliation{Fermi National Accelerator Laboratory, PO Box 500,
  Batavia, Illinois 60510, USA}

\author{R.~Bruce}
\author{S.~Redaelli}
\author{A.~Rossi}
\author{B.~Salvachua~Ferrando}
\affiliation{CERN, CH-1211 Geneva 23, Switzerland}

\date{30 October 2014}

\begin{abstract}
  Collimation with hollow electron beams is a technique for halo
  control in high-power hadron beams. It is based on an electron beam
  (possibly pulsed or modulated in intensity) guided by strong axial
  magnetic fields which overlaps with the circulating beam in a short
  section of the ring. The concept was tested experimentally at the
  Fermilab Tevatron collider using a hollow electron gun installed in
  one of the Tevatron electron lenses. Within the US LHC Accelerator
  Research Program (LARP) and the European FP7 HiLumi LHC Design
  Study, we are proposing a conceptual design for applying this
  technique to the Large Hadron Collider at CERN. A prototype hollow
  electron gun for the LHC was built and tested. The expected
  performance of the hollow electron beam collimator was based on
  Tevatron experiments and on numerical tracking simulations. Halo
  removal rates and enhancements of halo diffusivity were estimated as
  a function of beam and lattice parameters. Proton beam core
  lifetimes and emittance growth rates were checked to ensure that
  undesired effects were suppressed. Hardware specifications were
  based on the Tevatron devices and on preliminary engineering
  integration studies in the LHC machine. Required resources and a
  possible timeline were also outlined, together with a brief
  discussion of alternative halo-removal schemes and of other possible
  uses of electron lenses to improve the performance of the LHC.
\end{abstract}

\preprint{\parbox[c]{\textwidth}{CERN-ACC-2014-0248\hfill FERMILAB-TM-2572-APC}}

\maketitle

\clearpage
\tableofcontents
\clearpage

\section{Introduction}

Hollow electron beam collimation is a novel technique for beam
collimation and halo scraping~\cite{Shiltsev:HEBC:2007,
  Shiltsev:HEBC:2008}. It was tested experimentally at the Fermilab
Tevatron collider~\cite{Stancari:PRL:2011, Stancari:APSDPF:2011,
  Stancari:IPAC:2011, Stancari:CERN-Review:2012}. A magnetically
confined, possibly pulsed, low-energy (a few keV) electron beam with a
hollow current-density profile overlaps with the circulating beam over
a length of a few meters. If the electron distribution is axially
symmetric, the beam core is unperturbed, whereas the halo experiences
smooth and tunable nonlinear transverse kicks. The electron beam is
generated by a hollow cathode and transported by strong solenoidal
fields. The size, position, intensity, and time structure of the
electron beam can be controlled over a wide range of parameters.

The technique relies on robust conventional collimators to absorb
particles. However, it has several features that can complement a
classic multi-stage collimation system. In the case of high-power
proton beams, for instance, scraping is smooth, controllable, and the
issues of material damage are mitigated. A depletion zone is generated
between the proton beam core and the collimator edges, making local
energy deposition less sensitive to beam jitter, collimator movements,
orbit and tune adjustments, or fast failures in the case of
crab-cavity operation. It might be possible to reduce the
electromagnetic impedance of the conventional collimator jaws by
retracting them with respect to the standard configuration. Enhanced
halo diffusion and larger impact parameters may also improve the
overall cleaning efficiency; in the case of ions, these effects would
reduce uncontrolled losses due to fragmentation.

This method may provide a unique option to complement the LHC
collimation system. To study its implementation, a conceptual design
for the LHC upgrade was developed within the US LHC Accelerator
Research Program (LARP) and the European FP7 HiLumi LHC Design
Study. This may then develop into a technical design in 2014, with the
goal to build the devices in 2015--2017, after resuming LHC operations
and re-assessing needs and requirements with 6.5-TeV
protons. Installation during the next long LHC shutdown (LS2),
currently scheduled for 2018, would be technically possible. In case
of a resource-limited timeline, installation during the following long
shutdown (in 2022) is also an option. In this case, more advanced
solutions may be tested and included in the design.

\section{Motivation and strategy}

The requirements for improved beam collimation are being addressed
with high priority in preparation for the energy and high-luminosity
upgrades of the LHC. The present estimates are based on the
operational experience accumulated at 3.5~TeV and 4~TeV during the LHC
Run~1 and indicate that the halo cleaning performance of the present
collimation system is expected to be adequate for operations after the
current long shutdown (LS1)~\cite{LHCCollReview:2011,
  LHCCollReview:2013}. Caveats obviously apply due to the uncertainty
on the extrapolations to higher beam energies, intensities and
luminosities. A recent review of the LHC collimation project strongly
advised to study possible improvements of the present
system~\cite{LHCCollReview:2013}. While final decisions on further
upgrades can only be taken after sufficient operational experience at
higher energy, it is important to continue critical studies to
identify possible improvements for implementation in the next long
shutdown (LS2), starting in 2018. Hollow electron beam collimation is
considered as a promising option to enhance the present LHC
collimation.

In 2012, the primary collimator settings cut into the beam halo down
to 4.3\sigmap\ (where \sigmap\ is the rms proton beam size calculated for a
beam emittance of 3.5~$\mu$m), which was
required to push the amplitude function at the collision points
\betastar\ down to 60~cm~\cite{Bruce:Evian:2011}. This corresponded to
half gaps of about 1~mm, i.e. as small as the nominal design values
for 7-TeV operations. Under these conditions, and contrary to what was
observed in previous years with more relaxed collimator settings, the
operation was significantly affected by beam losses throughout the
operational cycle~\cite{Salvachua:Evian:2012}. About 40~fills were
lost due to various beam instabilities before establishing
collisions. The interplay between collimator impedance and beam-beam
effects is being investigated as a possible source of beam losses. The
outcome of a dedicated hollow electron lens
review~\cite{HEBCReview:2012} indicated that the functionality of the
hollow electron beams demonstrated at the Tevatron would be very
useful to improve the LHC operation is case of the beam losses
observed in 2012.

The present collimation system cannot easily be used for active and
smooth halo scraping during high-intensity operations. Scraping would
only be possible by intercepting halo particles with primary
collimator jaws, resulting in sharp loss spikes. The operation with
bulk material very close to the beam core poses also issues in terms
of collimator impedance and material robustness in case of failures,
which would not apply if electron beams were used.

It was therefore decided that hollow electron beam collimation studies
should be pursued with high priority~\cite{Redaelli:HLTC:2013}. The
immediate goal is to achieve a technical design report for the
construction of 2 hollow electron beam devices by 2015, when the needs
for beam scraping at the LHC can be addressed based on solid
operational experience at higher energy.

Although they are not the focus of this report, there are other
possible uses of electron lenses in the LHC: (a)~generation of tune
spread for Landau damping to stabilize the beams before collisions;
(b)~compensation of long-range beam-beam interactions in upgrade
scenarios with smaller crossing angles to improve luminosity, as an
alternative to compensation wires~\cite{Valishev:TM:2013}.

\section{Expected performance and parameter definitions}

\begin{table}
\caption{List of hollow electron lens
  parameters for the LHC. The requirements on the electron beam current
  stem from the magnetic rigidity of the proton beam,
  from the length of the interaction region, and from the size of the
  electron beam (Section~\ref{sec:hebc}). The size of the cathode is
  determined by the proton beam size, by the desired range of
  scraping positions, and by the available magnetic fields
  (Section~\ref{sec:geom}).}
\label{tab:par}
\begin{ruledtabular}
\begin{tabular}{lr}
Parameter & Value or range \\
\hline \multicolumn{2}{c}{\em Beam and lattice} \\ \hline
Proton kinetic energy, \Epkin\ [TeV] & 7 \\
Proton emittance (rms, normalized), \pemit\ [$\mu$m] & 3.75 \\
Amplitude function at electron lens, $\beta_{x, y}$ [m] & 200 \\
Dispersion at electron lens, $D_{x, y}$ [m] & $\leq 1$ \\
Proton beam size at electron lens, \sigmap\ [mm] & 0.32 \\
\hline \multicolumn{2}{c}{\em Geometry} \\ \hline
Length of the interaction region, \Lint\ [m] & 3 \\
Desired range of scraping positions, \rmi\ [\sigmap] & 4--8 \\
\hline \multicolumn{2}{c}{\em Magnetic fields} \\ \hline
Gun solenoid (resistive), \Bg\ [T] & 0.2--0.4 \\
Main solenoid (superconducting), \Bm\ [T] & 2--6 \\
Collector solenoid (resistive), \Bc\ [T] & 0.2--0.4 \\
Compression factor, $\kompr \equiv \sqrt{\Bm/\Bg}$ & 2.2--5.5 \\
\hline \multicolumn{2}{c}{\em Electron gun} \\ \hline
Inner cathode radius, \rgi\ [mm] & 6.75 \\
Outer cathode radius, \rgo\ [mm] & 12.7 \\
Gun perveance, \perv\ [$\mu$perv] & 5 \\
Peak yield at 10~kV, \Ie\ [A] & 5 \\
\hline \multicolumn{2}{c}{\em High-voltage modulator} \\ \hline
Cathode-anode voltage, \Vca\ [kV] & 10 \\
Rise time (10\%--90\%), \modrise\ [ns] &  200\\
Repetition rate, \modrep\ [kHz] & 35 \\
\end{tabular}
\end{ruledtabular}
\end{table}

In this Section, we describe the principles of hollow electron beam
collimation, its impact on beam halo dynamics, and the causes and
mitigation of possible unwanted effects on the beam core. A set of
working parameters (summarized in Table~\ref{tab:par}) is derived.

\subsection{Electron lenses and collimation with hollow
  electron beams}
\label{sec:hebc}

\begin{figure}
\includegraphics[width=0.75\textwidth]{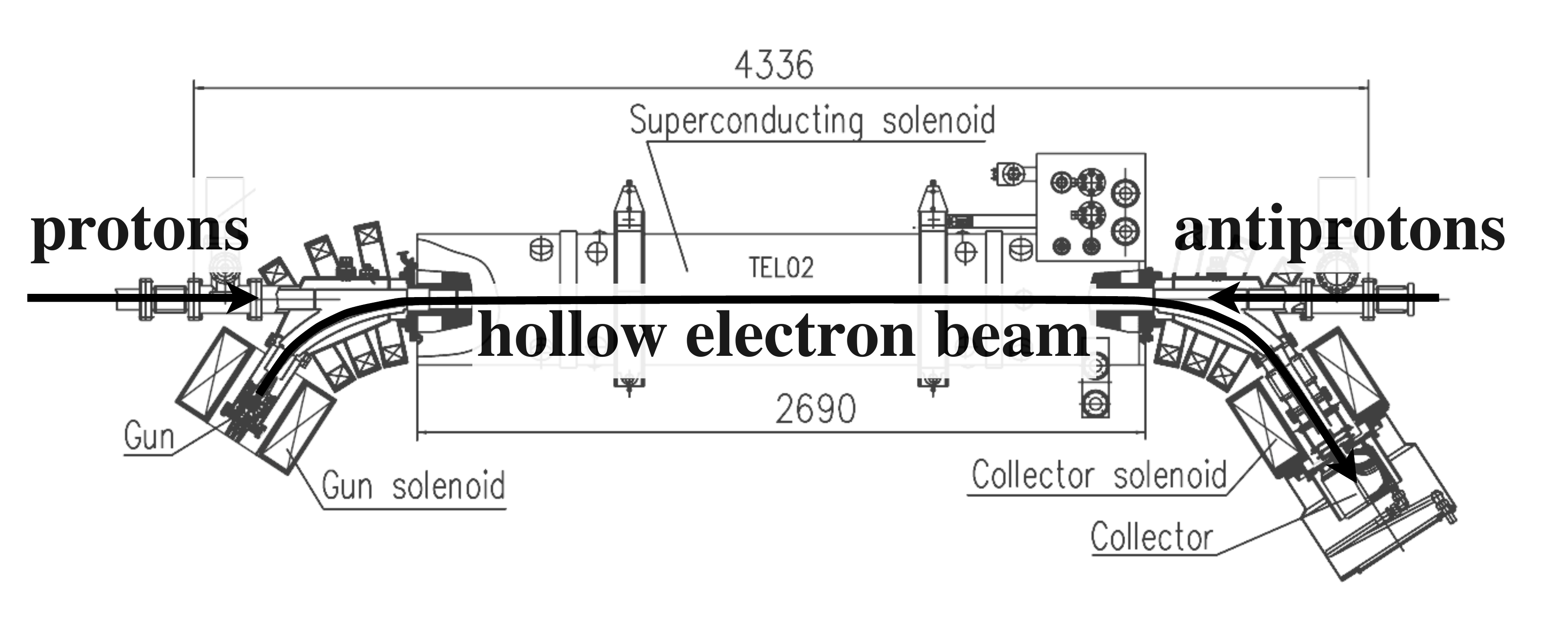}
\caption{Layout of the beams in the second Tevatron electron lens
  (TEL-2). The electron beam is generated and accelerated in the
  electron gun, transported through the overlap region with strong
  axial fields, and deposited in the collector. Dimensions are in
  millimeters.}
\label{fig:layout}
\end{figure}

\begin{figure}
\includegraphics[width=0.8\textwidth]{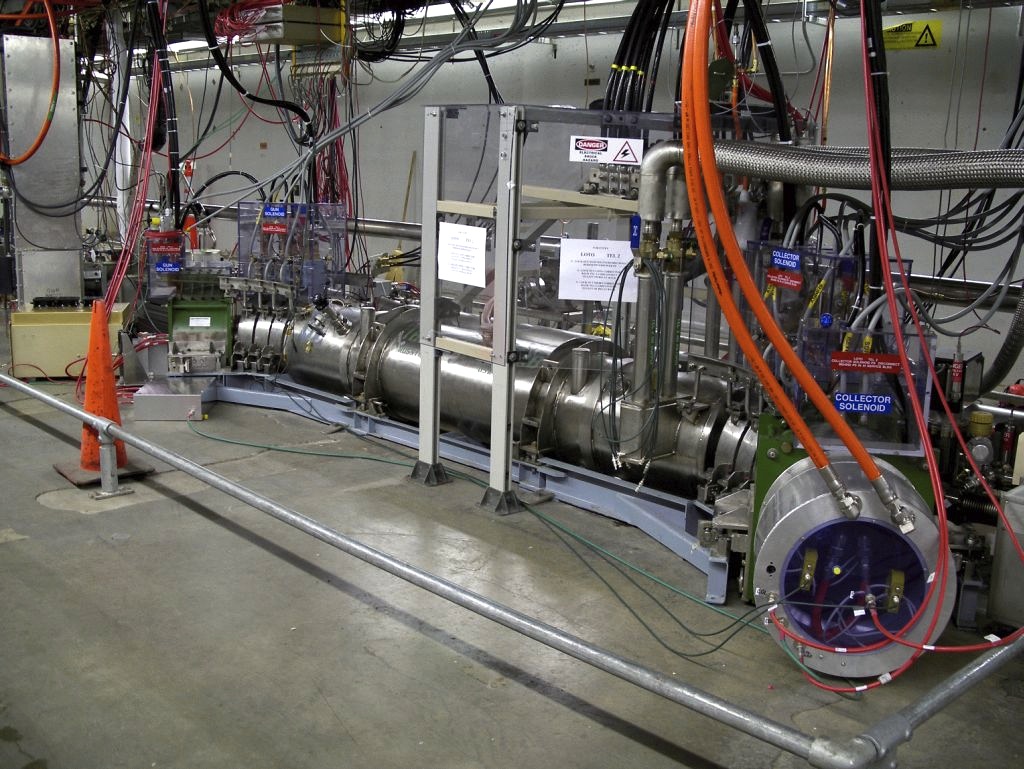}
\caption{Photograph of the second Tevatron electron lens (TEL-2) after
  installation in the Tevatron tunnel in 2006.}
\label{fig:TEL2-photo}
\end{figure}

\begin{figure}
\includegraphics[width=4in]{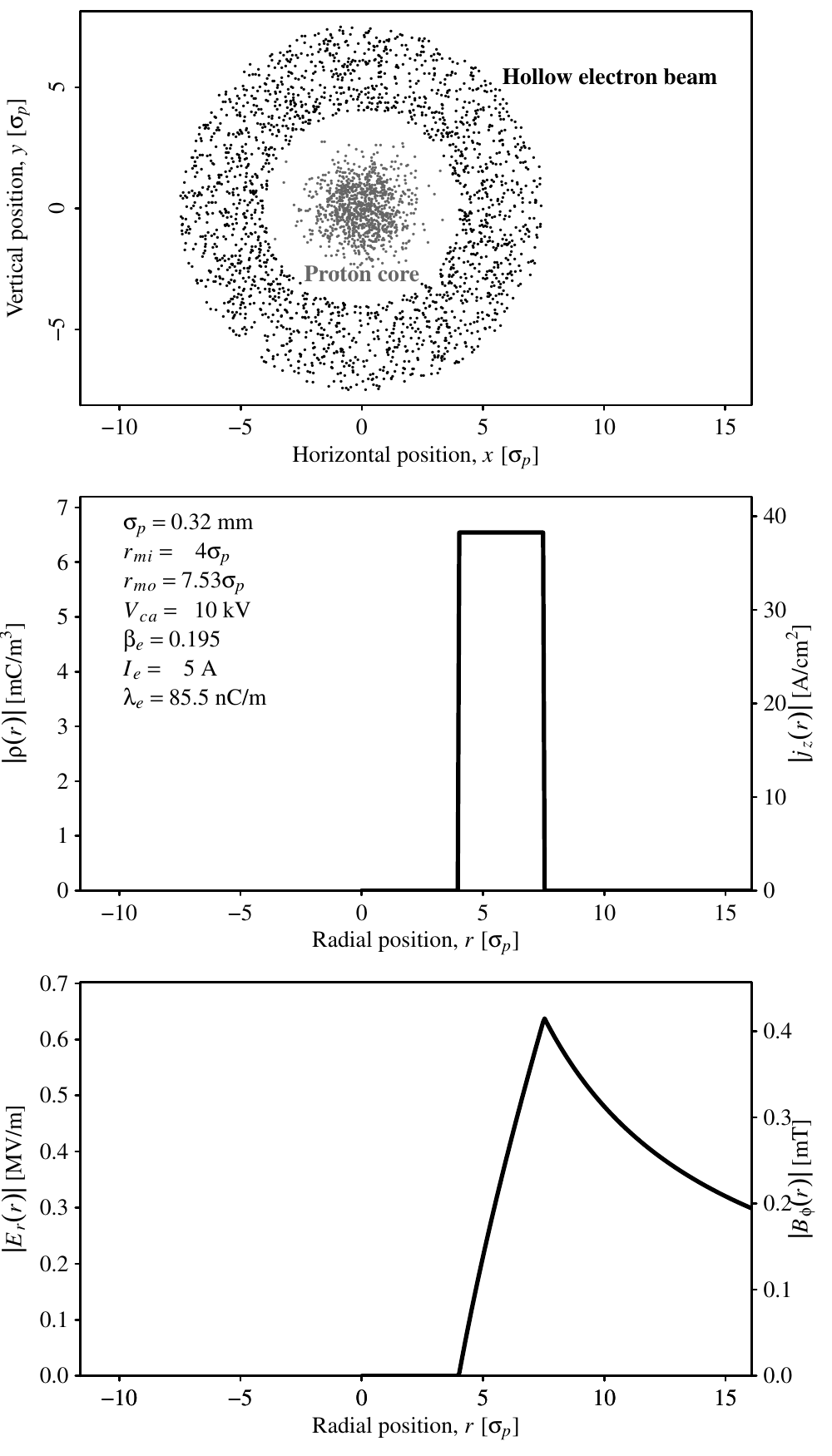}
\caption{Concept of hollow electron beam collimation. The top plot
  illustrates schematically the transverse layout of the beams in the
  overlap region assuming cylindrical symmetry. The bottom two plots
  show a numerical example: the electron charge density~$\rho$ and
  current density~$j_z$ as a function of radial position (middle
  plot); the radial electric field~$E_r(r)$ and azimuthal magnetic
  field~$B_\phi(r)$ generated by the electron beam (bottom plot).}
\label{fig:concept}
\end{figure}

Hollow electron beam collimation is based on the technology of
electron cooling and electron lenses. Electron lenses were developed
for beam-beam compensation in colliders~\cite{Shiltsev:PRSTAB:1999,
  Shiltsev:PRSTAB:2008, Shiltsev:NJP:2008}, enabling the first
observation of long-range beam-beam compensation effects by shifting
the betatron tunes of individual
bunches~\cite{Shiltsev:PRL:2007}. They were used for many years during
regular Tevatron collider operations for cleaning uncaptured particles
from the abort gap~\cite{Zhang:PRSTAB:2008}. Thanks to the reliability
of the hardware, one of the two Tevatron electron lenses (TEL-2) could
be used for experiments on head-on beam-beam compensation in
2009~\cite{Stancari:BB:2013}, and for exploring hollow electron beam
collimation in 2010--2011~\cite{Stancari:PRL:2011,
  Stancari:APSDPF:2011, Stancari:IPAC:2011}. Electron lenses for
beam-beam compensation were built for the Relativistic Heavy Ion
Collider (RHIC) at Brookhaven National Laboratory and are currently
being commissioned~\cite{Fischer:IPAC:2013}.

Figure~\ref{fig:layout} shows the layout of the beams in one of the
Tevatron electron lenses. The beam is formed in the electron gun
inside a conventional solenoid and guided by strong axial magnetic
fields. Inside the superconducting main solenoid, the circulating beam
interacts with the electric and magnetic fields generated by the
electrons. The electron beam is then extracted and deposited in the
collector.

The halo of the circulating beam, i.e.\ particles with betatron
amplitudes that exceed the inner radius of the hollow electron beam,
is affected by nonlinear transverse kicks
(Figure~\ref{fig:concept}). The angular kick~$\theta$ experienced by a
proton at radius~$r$ traversing a hollow electron beam enclosing
current~$I_{er}$ in an interaction region of length~\Lint\ is given by
the following expression:
\be
\theta = \frac{2 I_{er} \Lint (1\pm \beta_e \beta_p)}
              {r \beta_e \beta_p c^2 (B\rho)_p}
         \left(\frac{1}{4\pi \epsilon_0}\right),
\ee
where $v_e = \beta_e c$ is the electron velocity, $v_p = \beta_p c$
the proton velocity, and $(B\rho)_p$ is the magnetic rigidity of the
proton beam. The `$+$'~sign applies when the magnetic force is
directed like the electrostatic attraction ($\ve{v}_e \cdot \ve{v}_p <
0$), whereas the `$-$'~sign applies when $\ve{v}_e \cdot \ve{v}_p >
0$. For example, in a configuration with $I_{er}=\q{5}{A}$,
$\Lint=\q{3}{m}$, $\beta_e = 0.195$ (10-keV electrons), $r=\q{2.5}{mm}$,
the corresponding kick is $\theta = \q{0.3}{\mu rad}$ for 7-TeV
protons. Because of the betatron oscillations of the protons, the
transverse kicks have different magnitudes at each turn. The strength
of the kicks is proportional to the electron beam current and can be
easily controlled.  The particles in the core of the circulating beam
(whose amplitudes are smaller than the inner electron-beam radius) are
unaffected if the distribution of the electron charge is axially
symmetric.

The main advantages over conventional collimators are that the
transverse kicks are controllable, there is no material deformation or
damage, the magnetized hollow electron beam has a low impedance, and
the position and size of the electron beam are set by configuring the
magnetic-field transport.

The Tevatron experiments on hollow electron beam collimation were
conducted on antiprotons, mainly at the end of regular collider
stores. In some cases, the electron beam was turned on for the whole
duration of the fill after collisions were established. Because of the
flexible pulsing pattern of the high-voltage
modulator~\cite{Pfeffer:JINST:2011}, the electron beam could be
synchronized with a subset of bunches, providing a direct comparison
with the unaffected beam. The main results of hollow electron beam
collimation in the Tevatron can be summarized as
follows~\cite{Stancari:PRL:2011, Stancari:APSDPF:2011,
  Stancari:IPAC:2011, Stancari:CERN-Review:2012}:
\begin{itemize}
\item the use of the electron lens was compatible with collider
  operations during physics data taking;
\item the alignment of the electron beam with the circulating beam was
  accurate and reproducible;
\item the halo removal rates were controllable, smooth, and detectable;
\item with aligned beams and inner electron beam radii above
  5$\sigma$, there were no intensity or luminosity lifetime changes,
  or emittance growth in the core; below about 4.5$\sigma$, scraping
  started to occur, with observable effects on luminosity lifetime,
  but still no measurable core emittance growth;
\item loss spikes due to beam jitter and tune adjustments were
  suppressed;
\item the local effect of the electron beam on beam halo fluxes and
  diffusivities were directly measured with collimator scans.
\end{itemize}
In this report, we focus on the issues arising from the extension of
the technique to the Large Hadron Collider.

\subsection{Effects on halo dynamics}

\begin{figure}
\includegraphics[width=6in]{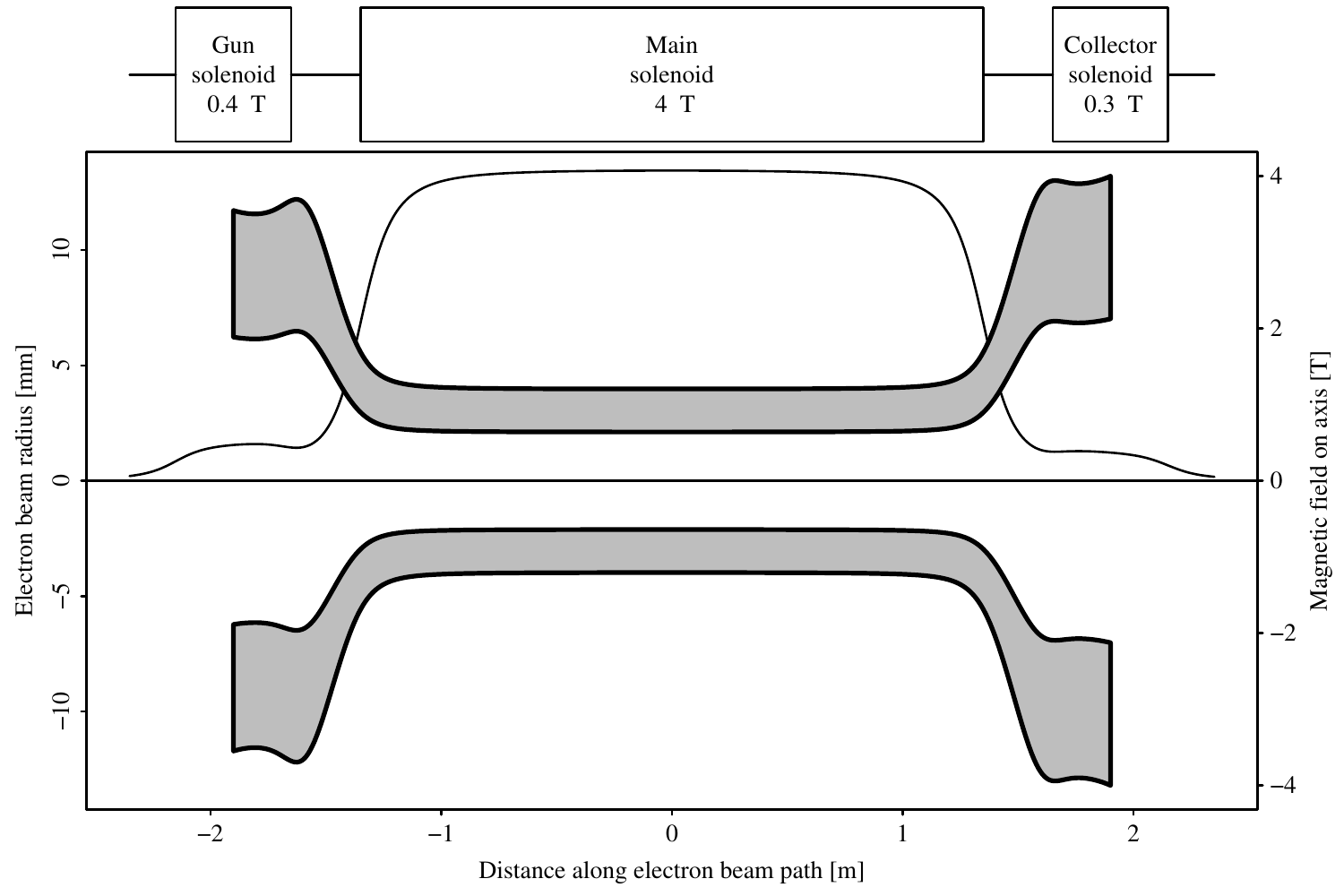}
\caption{Illustration of magnetic compression of the electron beam
  (gray) in an electron lens, as the axial magnetic field varies
  inside the solenoids (thin solid line).}
\label{fig:compression}
\end{figure}

\subsubsection{Beam optics and geometrical parameters}
\label{sec:geom}

The LHC primary collimators will be placed at around $6\sigmap$ from
the beam axis. For scraping the halo of a 7-TeV proton beam, we
envision the inner radius of the electron beam in the interaction
region~\rmi\ to be placed between about 4$\sigmap$ and 8$\sigmap$ of
the LHC proton rms beam size $\sigmap = \q{0.32}{mm}$. This size is
derived from the nominal normalized rms emittance $\pemit =
\q{3.75}{\mu m}$ and the typical amplitude function at the candidate
locations, $\beta = \q{200}{m}$. Scraping of elliptical proton beams
is possible with orbit bumps or by displacing the electron beam, but
for simplicity we focus on round beams.

For stability and for transport efficiency, the field in the guiding
solenoids should be as large as possible. Based upon previous
experience and technical feasibility, we consider configurations where
the gun, main (superconducting), and collector solenoids have fields
in the ranges
$\q{0.2}{T} \leq \Bg \leq \q{0.4}{T}$,
$\q{2}{T} \leq \Bm \leq \q{6}{T}$, and
$\q{0.2}{T} \leq \Bc \leq \q{0.4}{T}$,
respectively.  This implies magnetic compression factors~$\kompr \equiv
\sqrt{B_m/B_g}$ in the range
$2.2 \leq \kompr \leq 5.5$,
which sets the required sizes of the cathode inner and outer radii
(Figure~\ref{fig:compression}). The 1-inch electron gun cathode built
for this purpose (Section~\ref{sec:egun}), for instance, has inner
radius $\rgi = \q{6.75}{mm}$ and outer radius $\rgo =
\q{12.7}{mm}$. After magnetic compression, these radii translate to
$\q{1.2}{mm} = 3.9\sigmap \leq \rmi \leq 9.5\sigmap = \q{3.0}{mm}$ and
$\q{2.3}{mm} = 7.3\sigmap \leq \rmo \leq 18\sigmap = \q{5.7}{mm}$
in the interaction region inside the main solenoid, according to the
relation $\rgi^2 \cdot \Bg = \rmi^2 \cdot \Bm$ for magnetically
confined electron beams.

\subsubsection{Halo removal rates}

One of the main goals of the design study is to ensure that halo
removal rates for 7-TeV protons are detectable, usable, and
calculable.

The scraping experiments at the Tevatron with 0.98-TeV antiprotons
were done with peak electron beam currents up to 1.2~A. Halo removal
times ranged between seconds and minutes, depending upon the radius
and intensity of the electron beam. They were observable both with
colliding beams and with only antiprotons in the machine.

The transverse kicks generated by the hollow electron beam are
nonlinear and have a small random component due to noise in the
electron beam current. These kicks interact with the lattice
nonlinearities and with the sources of noise in the
machine. Therefore, the kicks needed to obtain a given halo removal
rate may not scale directly with the magnetic rigidity of the
circulating beam.

Tracking simulations in the Tevatron lattice with the \code{Lifetrac}
code showed that relatively small electron currents could
significantly enhance halo removal~\cite{Morozov:IPAC:2012}. The
removal rates are sensitive to the shape of the electron beam and to
the distribution of the halo population. It was observed that tracking
codes could give rough but conservative estimates of the removal
rates.  Numerical simulations of the LHC lattice with the
\code{SixTrack} code indicated that, in the absence of beam-beam
interactions and of diffusion processes, removal of 7-TeV protons with
a 1-A electron beam current would be slow~\cite{Previtali:IPAC:2013,
  Previtali:TM:2013}. These simulations were done with a simplified
halo distribution (horizontal only, no momentum spread) and without
collisions.

More realistic simulations with the \code{Lifetrac} code were
performed in the nominal LHC lattice (V6.503), with nominal beam
parameters, at 7~TeV, with and without
collisions~\cite{Valishev:TM:2014}. The machine lattice did not
include multipole errors. The hollow lens had the same nominal
parameters (1.2-A total current without turn-by-turn modulations,
inner radius at $4\sigmap$) and it was placed at the candidate
location in IR4 (see Section~\ref{sec:location}). The cleaning rate
for a uniform halo placed between $4\sigmap$ and $6\sigmap$ (Gaussian
in the longitudinal direction) was 2\% of the halo population per hour
without beam-beam interactions, and 30\% per hour with collisions.

The prototype LHC electron gun (Section~\ref{sec:egun}) had a yield of
over 5~A at 10~keV. This yield should be more than sufficient to have
a detectable effect on 7-TeV protons. At the yield of 3.6~A, the
simulations predict a halo cleaning rate of 40\% per hour without
collisions and up to 4\% per minute with collisions.

Different pulsing schemes were also pursued to extend the capabilities
of the technique, by exploiting the flexibility of the modulator
pulsing patterns. Most of the Tevatron scraping experiments were done
with the same turn-by-turn excitation intensity on the bunches of
interest. However, for beam-beam compensation purposes, the
high-voltage modulator was designed to handle bunch-by-bunch
adjustments, with 10\%--90\% rise times of
200~ns~\cite{Pfeffer:JINST:2011}. Moreover, fast abort-gap cleaning
was achieved by turning on the electron beam every 7th turn, in
resonance with the betatron oscillations of the uncaptured
beam~\cite{Zhang:PRSTAB:2008}.

In the LHC, one could change the electron beam current turn by turn,
synchronizing the voltage change with the abort gap, for
instance. Train-by-train (900-ns separation) or even batch-by-batch
(225~ns) intensity modulations are feasible; this allows one to
preserve the halo on a subset of bunches for diagnostics and machine
protection. Bunch-by-bunch adjustments every 25~ns or 50~ns would be
challenging and are probably unnecessary.

This flexibility opens up the possibility to operate the hollow
electron lens in different pulsing modes:
\begin{itemize}
\item \emph{continuous} --- the same voltage is applied every turn;
\item \emph{resonant} --- the voltage is changed turn by turn
  according to a sinusoidal function (possibly including a frequency
  sweep to cover the tune spread of the halo), or with the same
  amplitude, but skipping a given number of turns (as in the Tevatron
  abort-gap cleaning mode);
\item \emph{stochastic} --- the voltage is turned on or off every turn
  according to a random function, or a random component is added to a
  constant voltage amplitude.
\end{itemize}
These modes of operation were simulated with tracking
codes~\cite{Previtali:IPAC:2013, Previtali:TM:2013,
  Valishev:TM:2014}. Both the resonant and the stochastic mode gave
significant and tunable halo removal rates. While the first was
sensitive to the details of the tune distribution (lattice
nonlinearities, beam-beam interactions), the stochastic mode was much
more robust.

The introduction of stochastic turn-by-turn modulation of the electron
beam current significantly ehnances the halo cleaning efficiency,
making the electron lens the dominant loss-driving mechanism. The
cleaning rates for the cases with and without beam-beam interactions
do not differ as much as in the continuous mode. In either case, 50\%
of halo is removed in 200~s with a yield of 1.2~A, and 80\% at
3.6~A. The maximum cleaning rate attained in the stochastic mode was
about 100\% per minute.

\subsubsection{Diffusion enhancement}

Using collimator scans, it was possible to measure the effects of
collisions and of the hollow electron lens on halo diffusion in the
Tevatron as a function of betatron amplitude~\cite{Stancari:HB:2012,
  Stancari:BB:2013b}. The hollow electron lens could enhance halo
diffusivity in action space by two orders of magnitude. Diffusivities
in action space with and without collisions were also measured in the
LHC~\cite{Valentino:PRSTAB:2013}. Halo suppression is the main focus
of this project and the main consequence of the drift and diffusion
enhancement by the electron beam. However, we intend to further
investigate other aspects as well, such as the increase in impact
depth on the primary collimators and the possible resulting
improvement of collimation efficiency.

\subsubsection{Other effects of halo depletion}

Particle removal was not the only effect that could be measured in the
Tevatron. Thanks to the gated loss monitors (Section~\ref{sec:hebc}),
other consequences of halo depletion could be
observed~\cite{Stancari:APSDPF:2011, Stancari:IPAC:2011,
  Stancari:CERN-Review:2012}: the suppression of Fourier components of
losses related to beam jitter; the removal of the correlations between
losses from different bunch trains due to orbit fluctuations; and the
suppression of loss spikes induced by collimator setup or by tune
adjustments. Because of the much larger beam power in the LHC, the
capability to distribute losses in time may prove very useful.

\subsection{Undesired effects on the core}

\subsubsection{Current-density asymmetries in the electron beam}

The core of the circulating beam is unaffected if the distribution of
the electron charge is axially symmetric. One possible cause of
asymmetry is the space-charge evolution of the electron beam. Other
sources of asymmetry are the bends that are used to inject and extract
the electron beam from the interaction region.

The electron beam was turned on for several hours during some Tevatron
collider stores. With aligned beams and continuous operation, no
deterioration of the core lifetimes, emittances, or luminosities were
observed. Only a limited number of experiments were done in resonant
mode (by skipping turns). In these cases, the electron lens caused
emittance growth and luminosity degradation. A quantitative analysis
of the experiments is under way.

The current-density profiles generated by the hollow electron guns
were measured in the Fermilab electron-lens test stand as a function
of beam current and axial magnetic field. Space-charge evolution of
the electron beam profiles was mitigated by increasing the guiding
magnetic fields. Experiments in the test stand, analytical
calculations, and numerical simulations with the \code{Warp}
particle-in-cell code~\cite{Vay:CSD:2012} confirmed that, for main
fields above 2~T and beam currents up to several amperes, transverse
current-density profiles were practically frozen.

The calculation of the electric fields from the measured current
density profiles and the generation of the kick maps caused by the
bends is described in Refs.~\cite{Stancari:PAC:2013,
  Stancari:FN:2014}. These fields were used as inputs for tracking
simulations to estimate beam lifetimes and emittance growth rates. For
the Tevatron lattice and working point, the only azimuthal asymmetry
seen to cause extra losses in the core was the quadrupole component in
a particular resonant mode (pulsing every 6th
turn)~\cite{Morozov:IPAC:2012}. In LHC simulations with
\code{Lifetrac}, the bends in continuous mode had no effect on
lifetimes, emittances, or dynamic aperture~\cite{Valishev:TM:2014}.
However, the simulations suggest that, for the stochastic mode, the
uncompensated dipole component of the bending section kick may
introduce emittance growth that depends on the electron lens
design. Namely, the gun-side and collector-side bending sections of
the electron lens can be either on the same side of the device (as in
the Tevatron and RHIC electron lenses), or on opposite sides of the
device with respect to the beam propagation. In the former case, the
dipole components of horizontal kick from the bends add up, which
leads to the horizontal emittance growth. In the latter case, the
dipole components subtract leaving only higher order multipole
harmonics. The impact of these higher order harmonics on luminosity
lifetime is estimated at about 1\% per hour. Although this effect is
undesirable, it is slow compared to the scraping time scales
envisioned for the stochastic mode. Moreover, further optimizations of
the bending sections are possible. Because the stochastic mode of
operation offers greater flexibility, the above considerations point
towards an electron lens design with the gun and the collector bends
on opposite sides.

\subsubsection{Impedance of the electron beam}
\label{sec:e-beam-imped}

An early concern on the use of electron lenses for beam-beam
compensation in colliders was the stability of the beams. The electron
beam is continuously renewed, so only intrabunch effects were
important in the Tevatron. In particular, a displaced head of the
circulating bunch could distort the electron beam, whose
electromagnetic fields could in turn act back on the bunch tail,
causing oscillations in the electron trajectory and a fast transverse
mode coupling instability. A 10-keV electron beam traverses the
overlap region of 3~m in about 50~ns. For LHC bunch spacings of 25~ns
or 50~ns, coupled-bunch modes may need to be included.

The electron beam is made stiff by increasing the axial solenoidal
field, reducing its effective impedance. Instability thresholds for
the head-on beam-beam case were estimated in
Ref.~\cite{Burov:PRE:1999}. The stability of the system was indirectly
confirmed by routinely operating the Tevatron electron lenses above
1~T. For the hollow-beam case, requirements are expected to be much
less stringent because of the smaller fields generated by a distorted
hollow density distribution near its axis. The impedance of the
electron-lens hardware (without electron beam) is discussed in
Section~\ref{sec:imped}.

\subsection{Further experimental tests}

Electron lenses for head-on beam-beam compensation are being
commissioned at RHIC~\cite{Fischer:IPAC:2013}. It was suggested that
further experiments with hollow electron beams on protons for the LHC
could address some of the operational scenarios not tested at the
Tevatron, such as dynamical use during ramp and squeeze, or a
systematic study of pulsed modes.

Although appealing, this option does not appear very likely due to the
priorities and beam availability at BNL. Obviously, the first priority
is to commission the electron lenses for beam-beam compensation. The
2014 run will focus on ion operations with Au-Au collisions. The
earliest operation with protons ($p$-$p$ or $p$-Au) is currently
scheduled for 2015.

\section{Hardware specifications and integration studies}

In this Section, we describe some of the practical aspects of the
implementation of electron lenses in the LHC, taking into account what
was achieved with the Tevatron and RHIC electron lenses and the
specific LHC conditions.  This work will serve as the basis for a
detailed technical design report. Table~\ref{tab:par} summarizes the
main characteristics of the device. A detailed description of the
Tevatron hardware can be found in Ref.~\cite{Shiltsev:PRSTAB:2008}.

\begin{figure}
\includegraphics[width=0.8\textwidth]{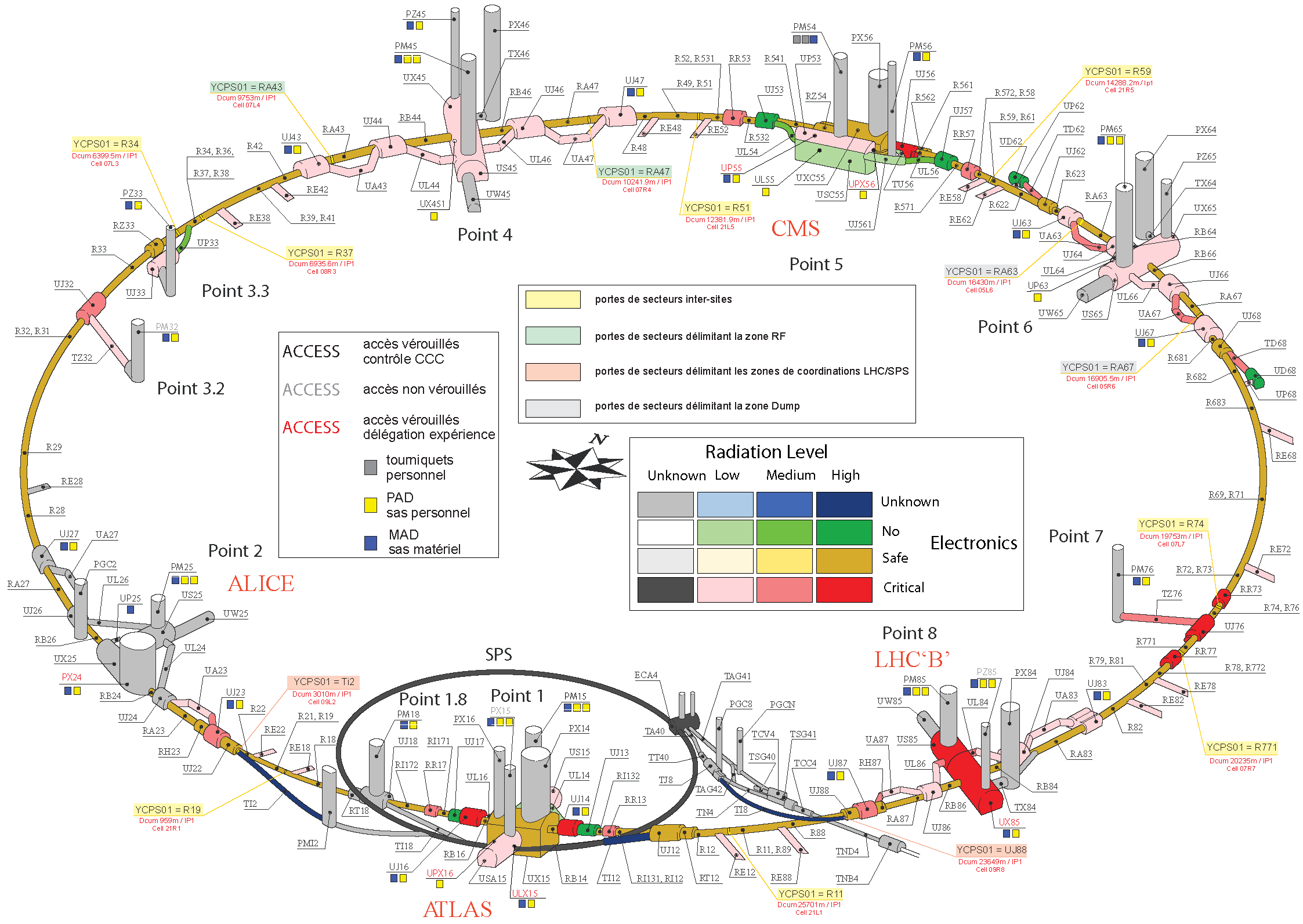}
\caption{Schematic diagram of the LHC. Candidate locations for the
  electron lenses are RB-44 and RB-46 at Point~4, on each side of the
  interaction region IR4, which houses the accelerating cavities.}
\label{fig:loc-schematic}
\end{figure}

\begin{figure}
\includegraphics[width=0.8\textwidth]{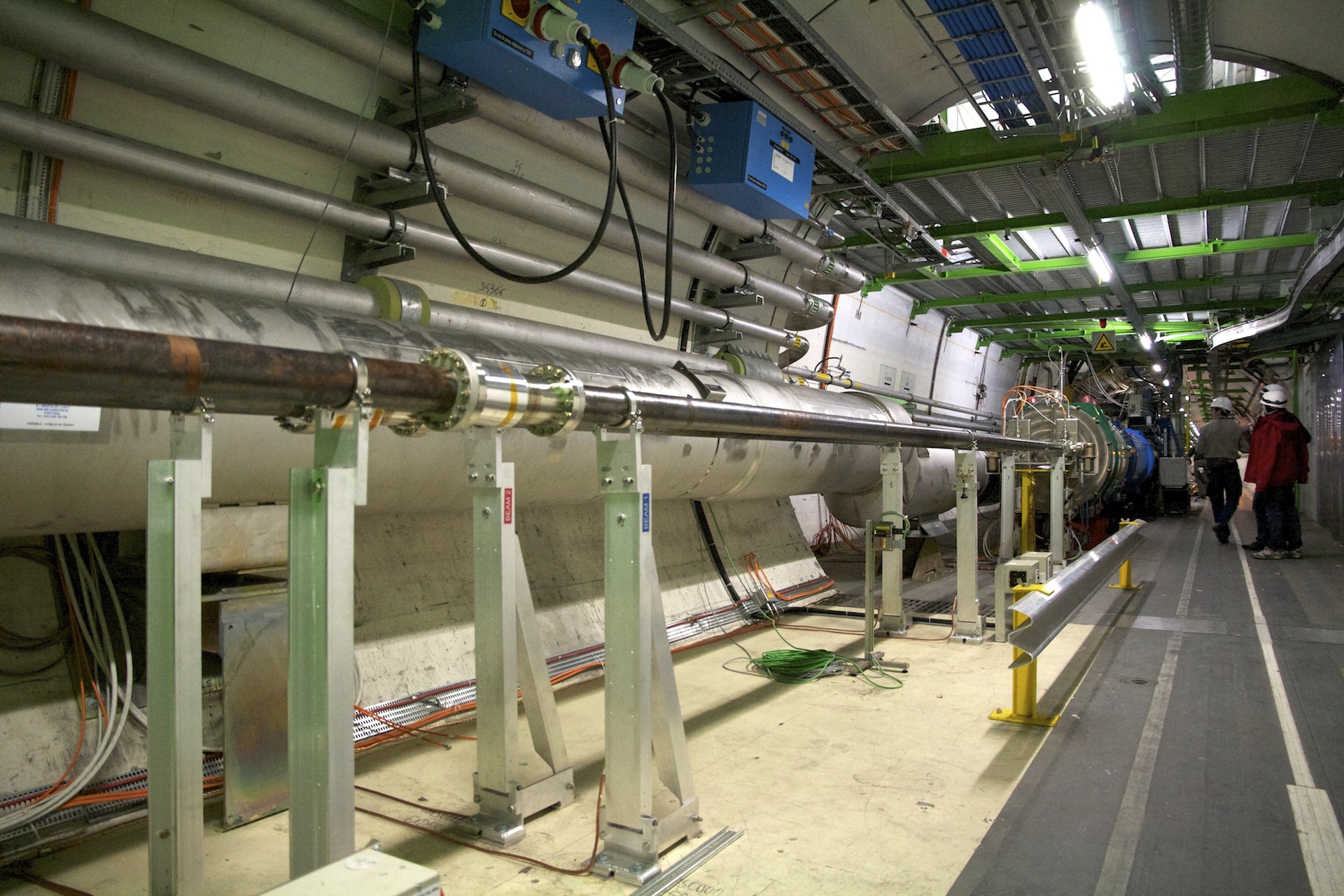}
\caption{Photograph of RB-46, one of the candidate locations, east of
  IR4. In this view, Beam~1 is on the inside, moving away from the
  viewer. The first downstream element is the green synchrotron-light
  undulator. The interaxis beam-pipe separation is 420~mm. The RB-44
  location has a very similar (mirror-imaged) configuration. (Photo
  taken by V.~Previtali on November~10, 2011.)}
\label{fig:loc-photo}
\end{figure}

\begin{figure}
\includegraphics[width=0.49\textwidth]{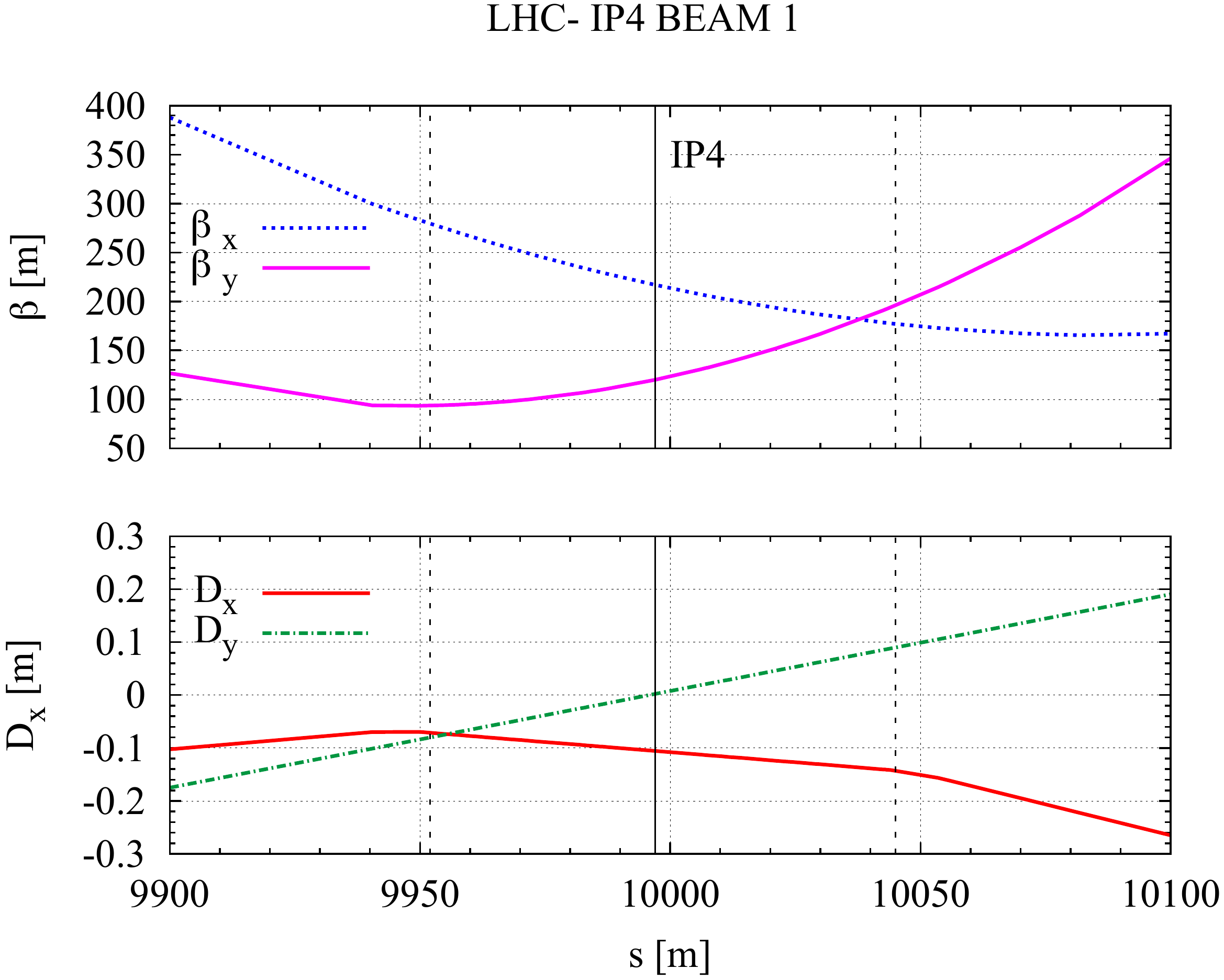}
\hfill
\includegraphics[width=0.49\textwidth]{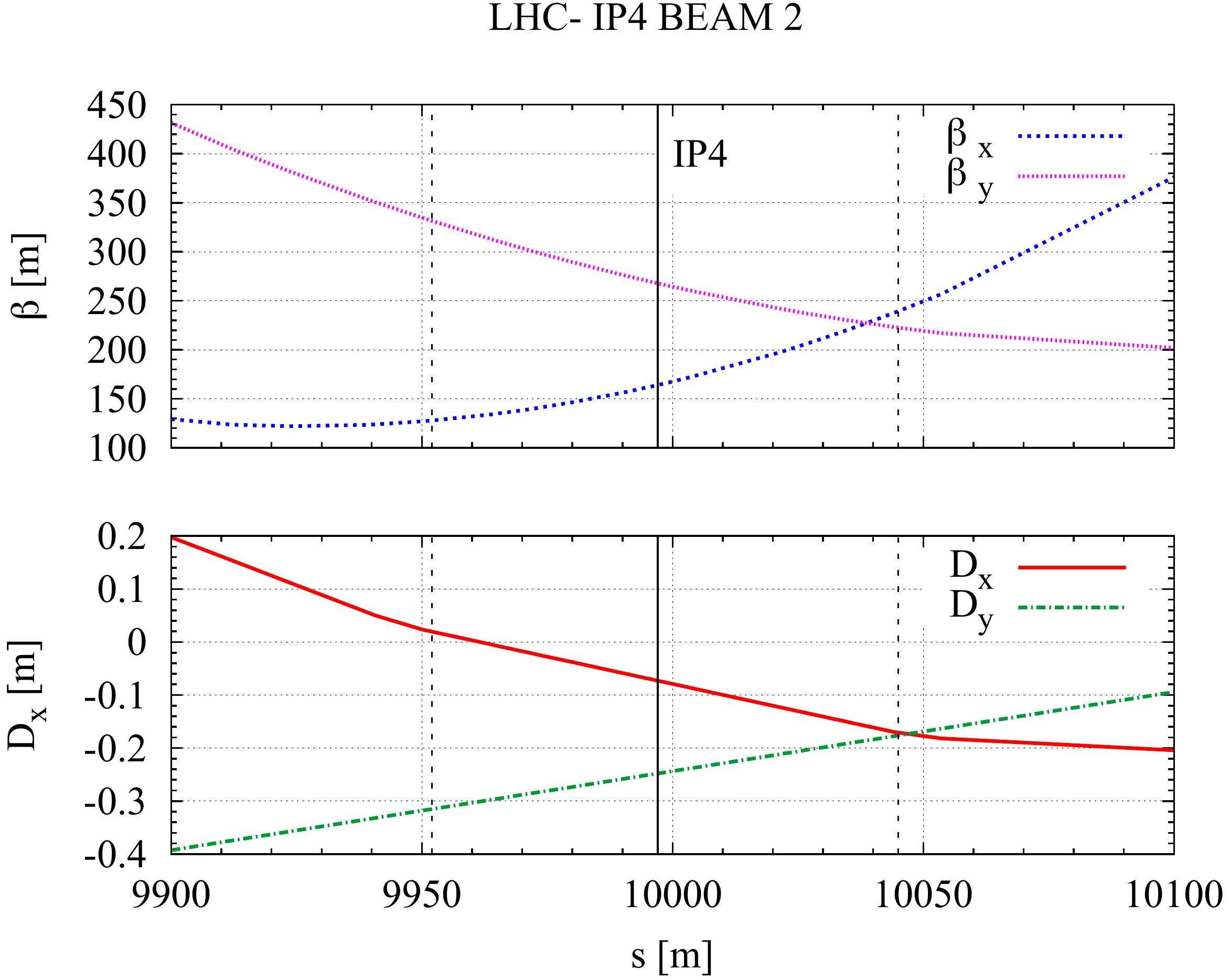}
\caption{LHC machine lattice near the interaction region IR4. The
  candidate locations RB-44 (smaller $s$ coordinate) and RB-46 (larger
  $s$) are marked with the dashed lines.}
\label{fig:lattice}
\end{figure}

\begin{figure}
\includegraphics[height=0.38\textwidth]{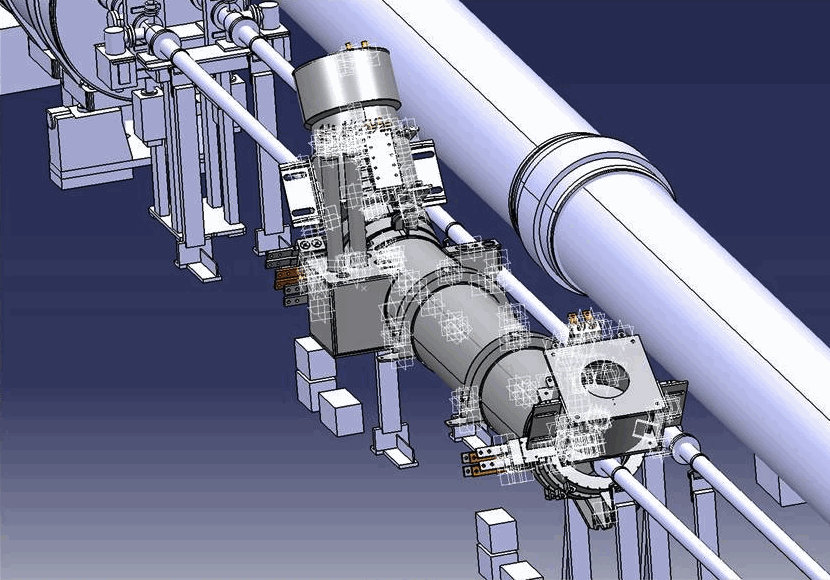}
\hfill
\includegraphics[height=0.38\textwidth]{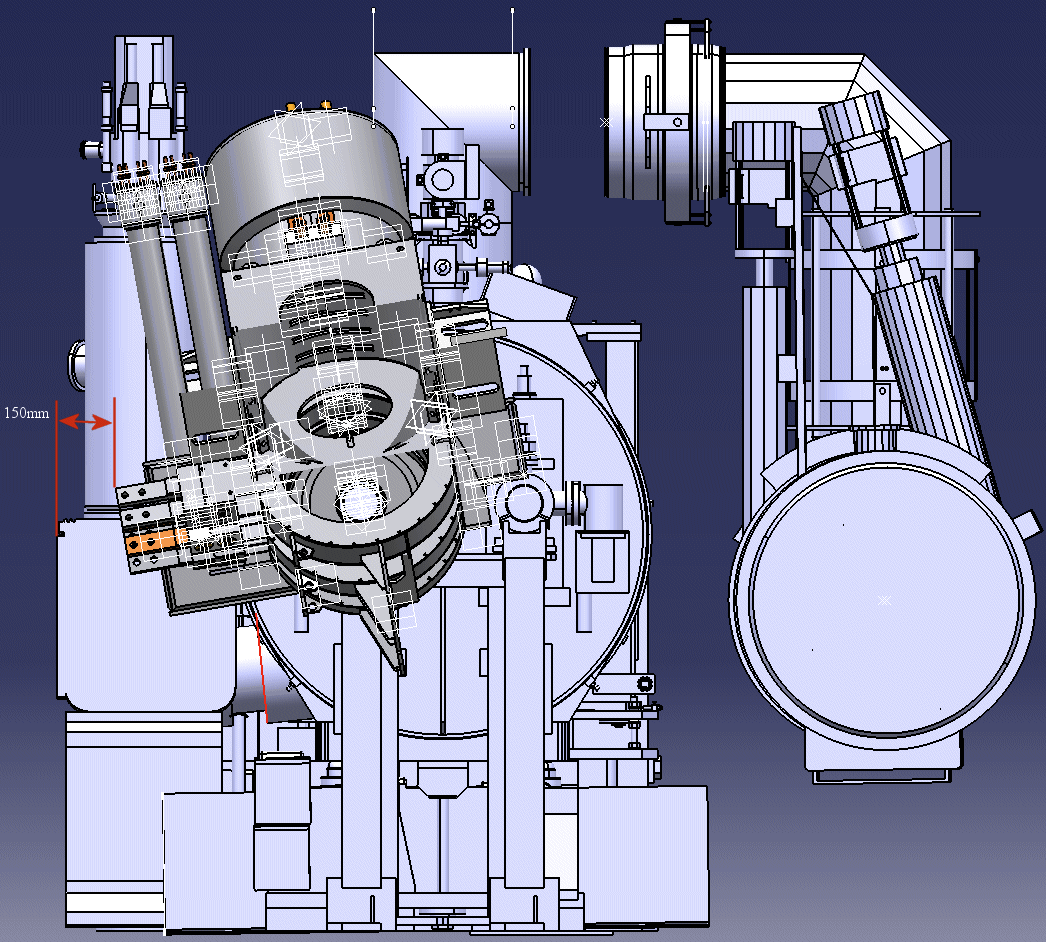}
\caption{Integration study of a Tevatron electron lens (TEL-2) at the
  RB-44 location in LHC. Transverse space constraints require a
  rotation of 80$^\circ$ around the beam axis with respect to the
  Tevatron configuration.}
\label{fig:integration}
\end{figure}

\begin{figure}
\includegraphics[width=0.24\textwidth]{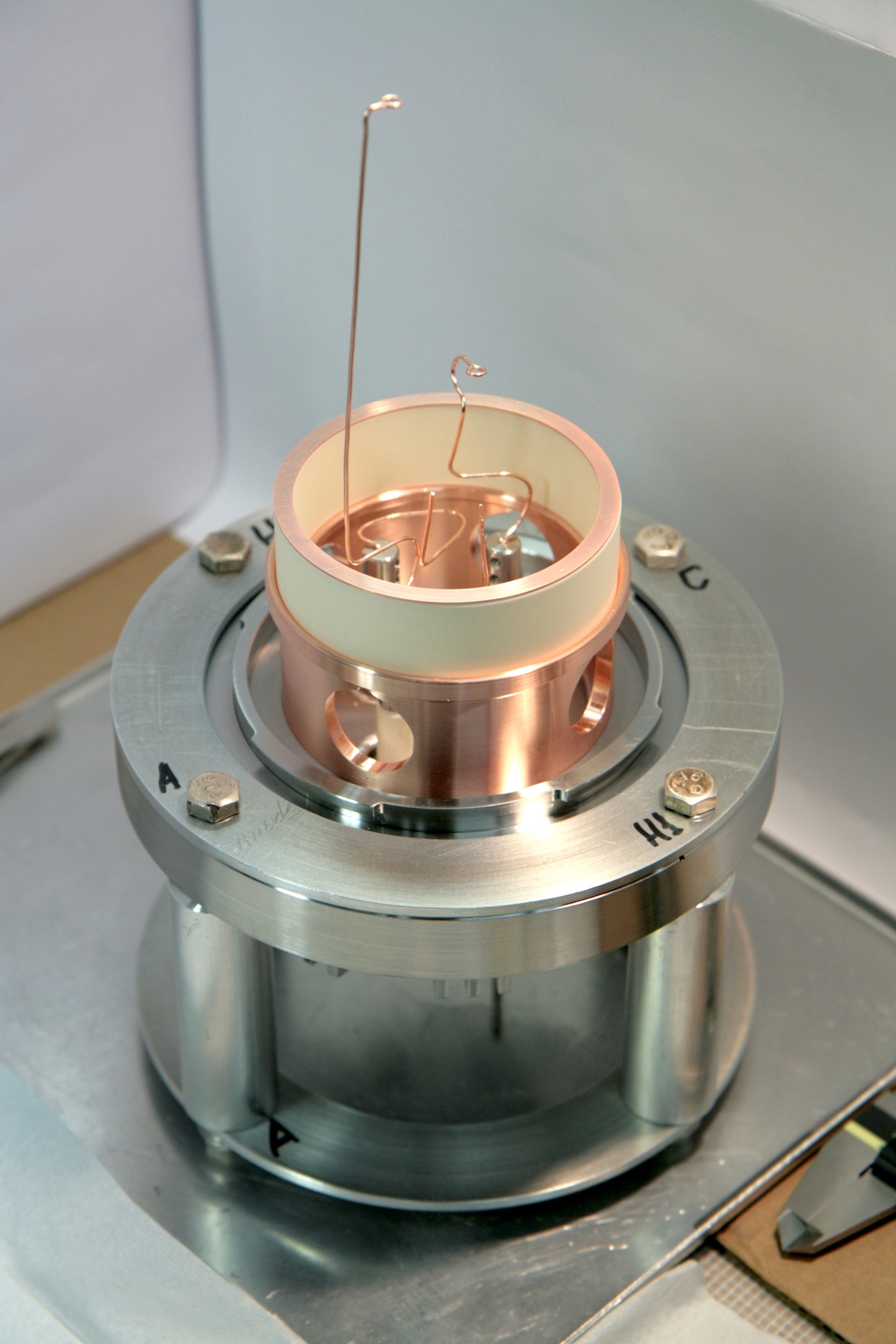}
\includegraphics[width=0.24\textwidth]{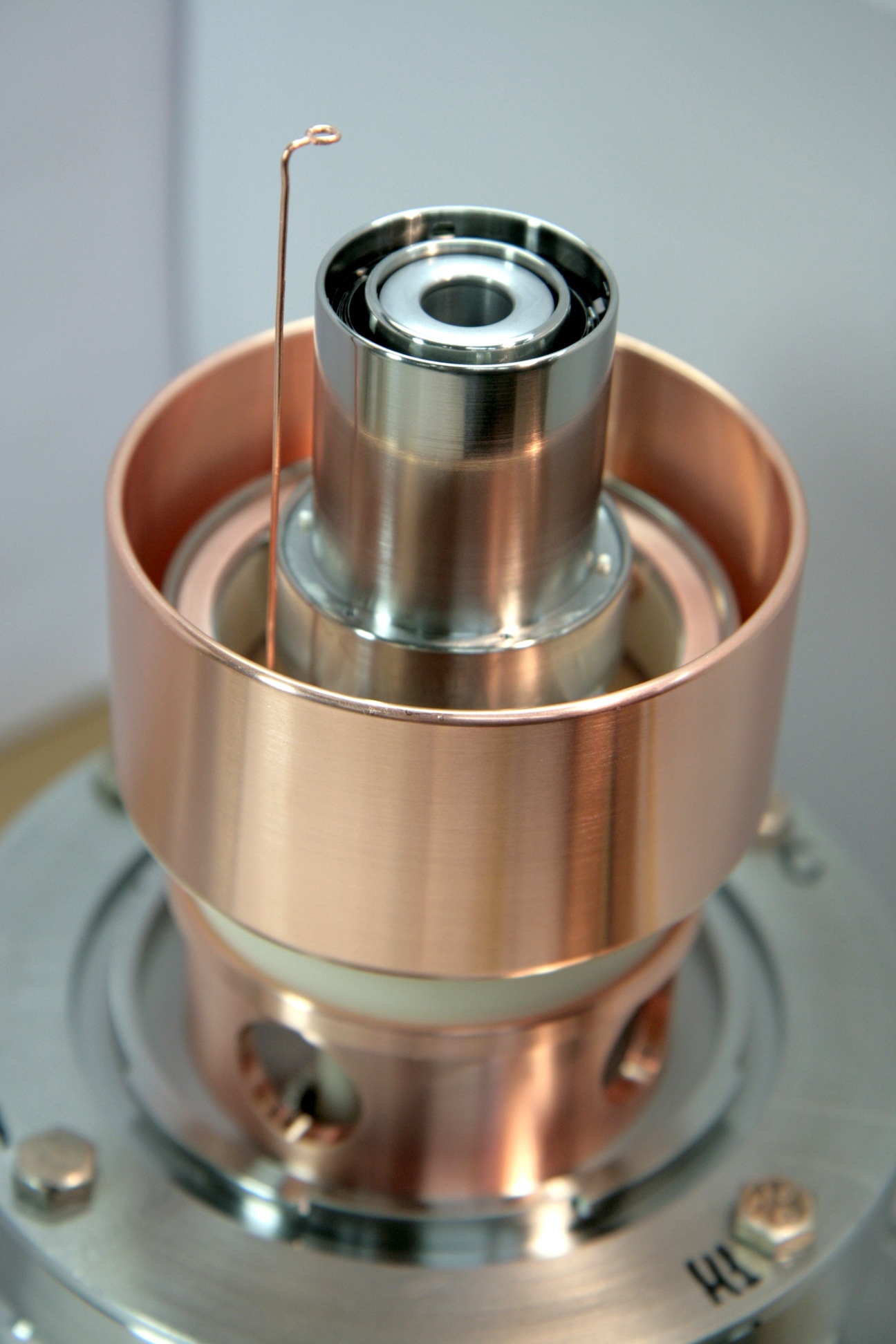}
\includegraphics[width=0.24\textwidth]{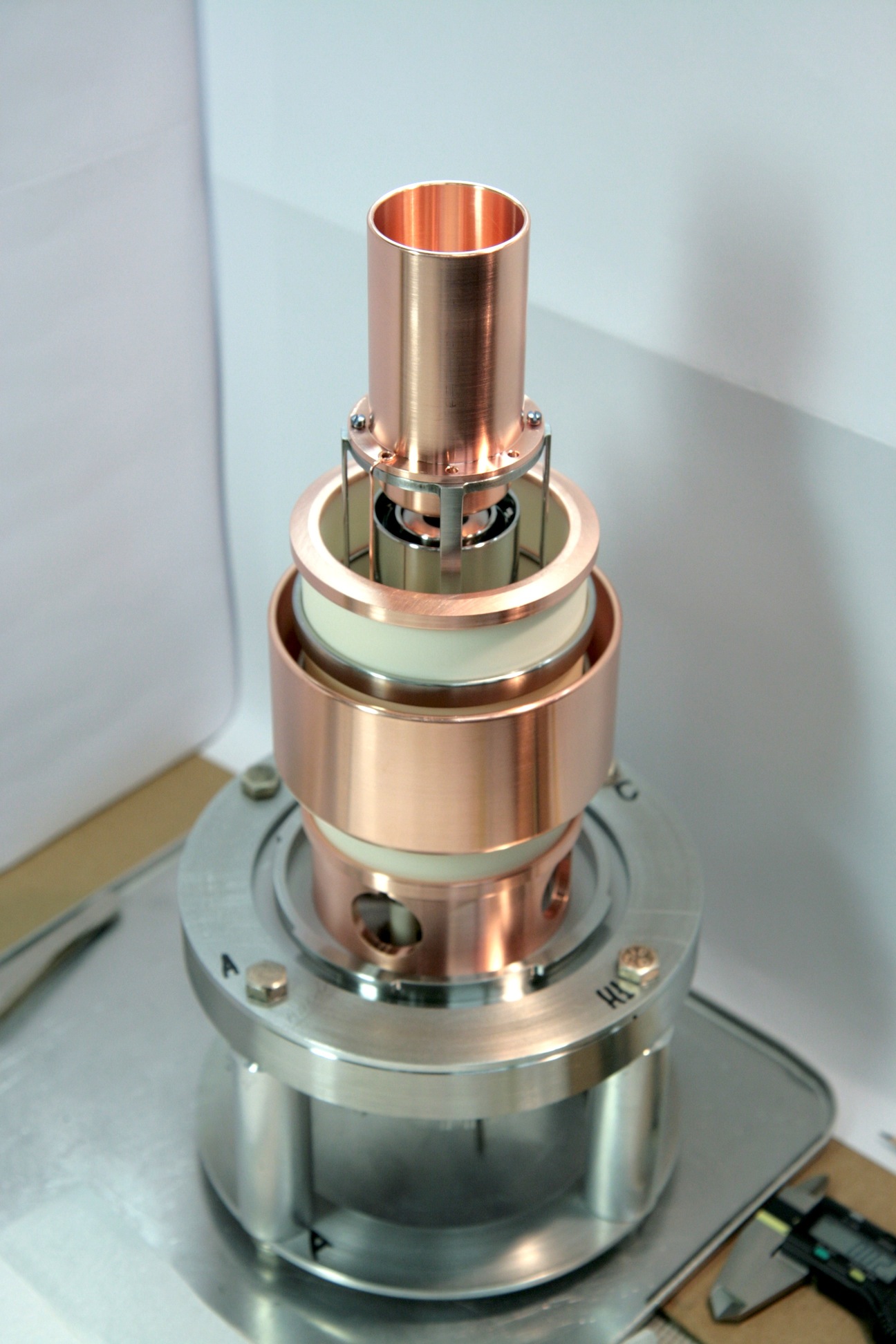}
\includegraphics[width=0.24\textwidth]{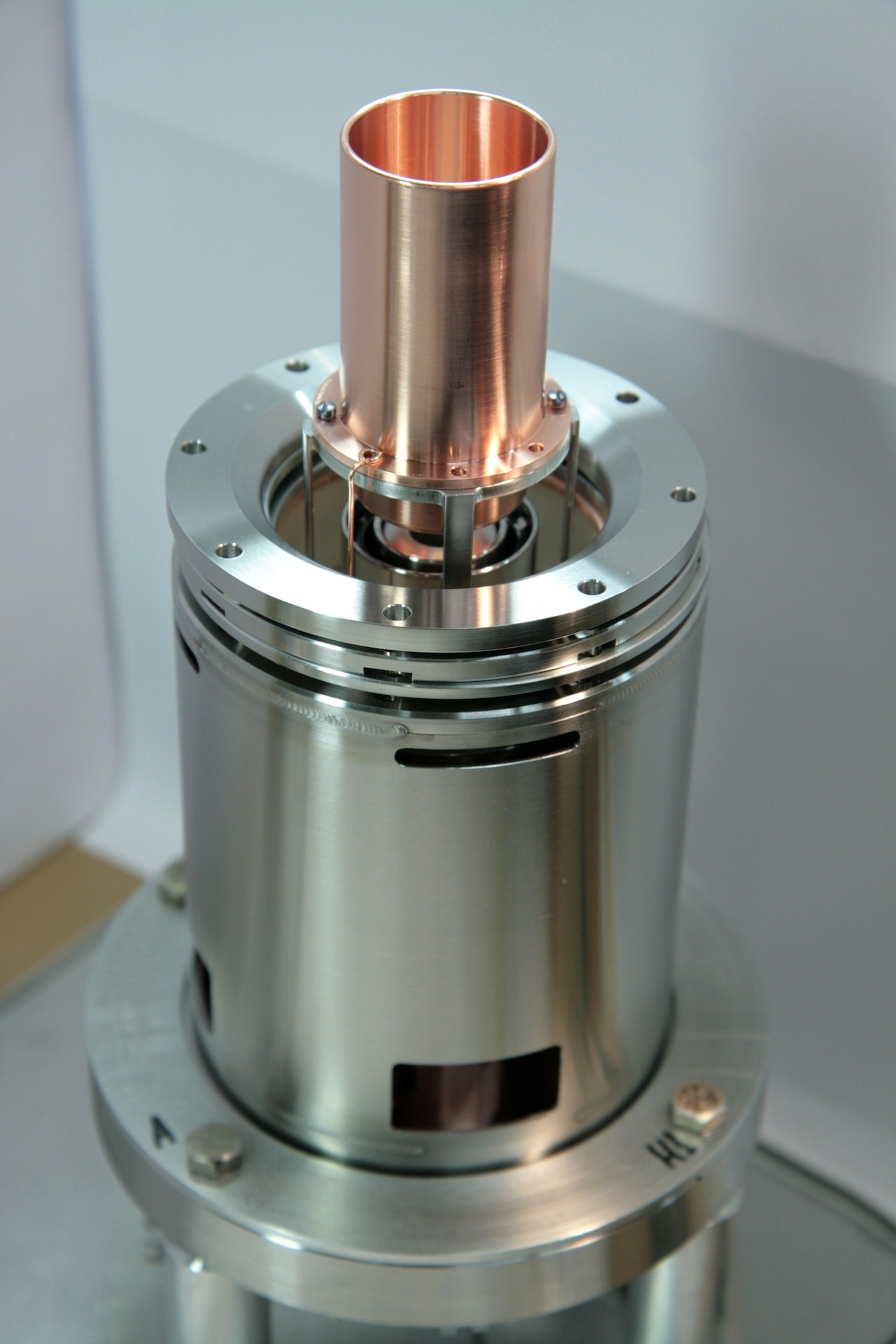}
\caption{Assembly of the prototype (1-inch) hollow electron gun. The
  first photograph shows the base flange with electrical
  connections. In the second photo, one can see the hollow cathode
  with convex surface and the rim of the control electrode; both are
  surrounded by cylindrical heat shields. The mounting of the copper
  anode is shown in the third picture. The last picture shows the
  complete assembly.}
\label{fig:egun-assembly}
\end{figure}

\begin{figure}
\includegraphics[width=3.25in]{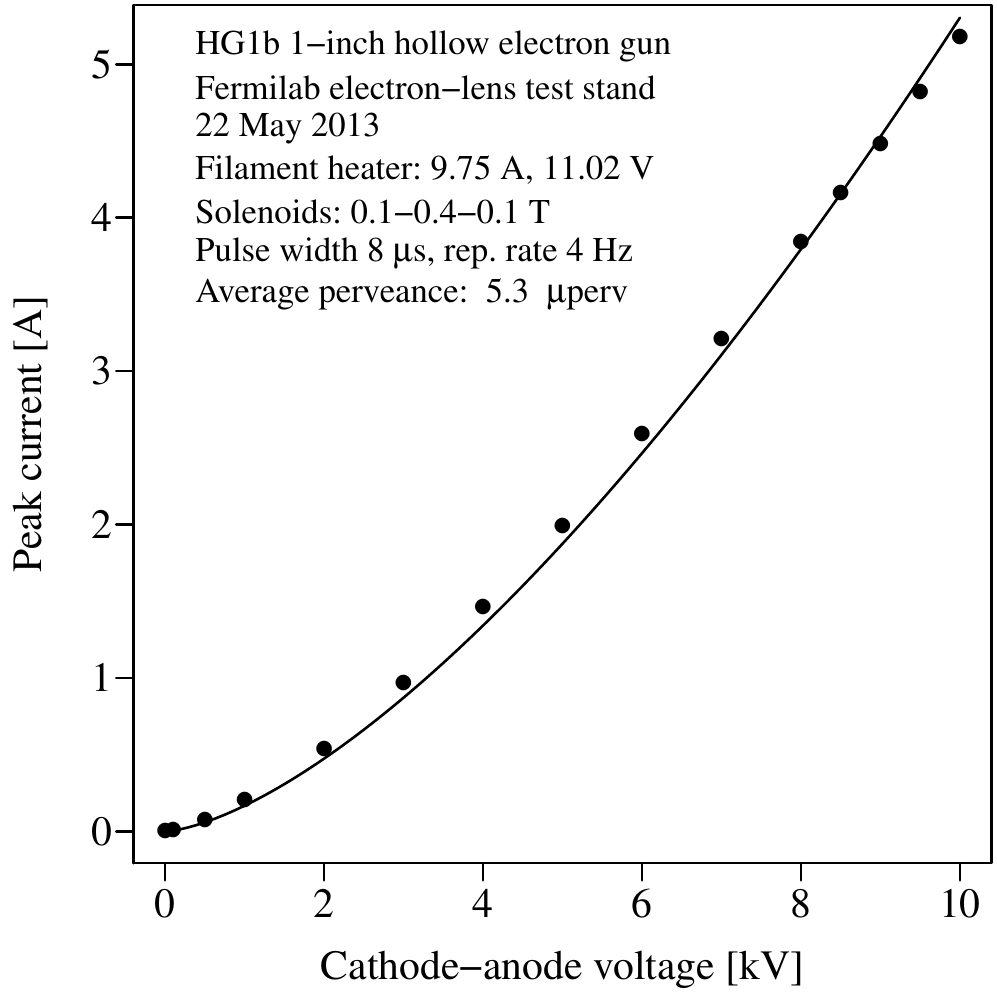}
\caption{Performance of the prototype (1-inch) hollow electron gun
  measured at the Fermilab electron-lens test stand. The total peak
  current at the cathode~\Ie\ is plotted as a function of the
  cathode-anode voltage~\Vca.}
\label{fig:egun-performance}
\end{figure}

\subsection{Physical and mechanical features}
\label{sec:location}

The second Tevatron electron lens (TEL-2) occupies 5.8~m of tunnel
length, is 1.7-m wide, 1.5-m tall (including current and cryogenic
leads), and weighs about 2~t. The radius of the cryostat is 0.3~m. We
first considered reusing TEL-2 and installing it in the LHC. Candidate
locations (RB-44 and RB-46) were identified on each side of the
radiofrequency insertion at IR4 (Figures~\ref{fig:loc-schematic}
and~\ref{fig:loc-photo}). In addition to the available longitudinal
space, these locations were originally chosen because of the
availability of cryogenic infrastracture and because of the large
interaxis distance (420~mm) between the two beam pipes to accomodate
the TEL-2 cryostat. Beam optics is also favorable
(Figure~\ref{fig:lattice}): the beams are practically round and the
lattice functions are of the order of 200~m.  Three-dimensional
drawings of TEL-2 were produced. Preliminary integration studies by
Y.~Muttoni's team showed that the hardware would fit, but it would
require a rotation of the cryostat and of the gun/collector solenoids
(Figure~\ref{fig:integration}). Although this is feasible, the design
of new cryostats for the LHC tunnel would probably be preferable.

\subsection{Hollow electron guns}
\label{sec:egun}

A prototype hollow electron gun for the LHC was designed, built, and
tested at the Fermilab electron-lens test stand
(Figure~\ref{fig:egun-assembly}). Its design was based on previous
electron guns used in the Tevatron. The tungsten dispenser cathode
with BaO:CaO:Al$_2$O$_3$ impregnant has an annular shape and a convex
surface to increase perveance~\cite{Sharapa:NIMA:1998}. The outer
diameter is 25.4~mm and the inner diameter is 13.5~mm. A filament
heater was used to reach the operating temperature of 1400~K. The
shape of the extraction electrodes to achieve the desired
current-density distribution in the space-charge-limited regime were
calculated with the \code{UltraSAM} code~\cite{Ivanov:EPAC:2002}. This
gun had a perveance of \q{5.3}{\mu perv}. This means that it could
yield more than 5~A of peak current at a cathode-anode voltage of
10~kV (Figure~\ref{fig:egun-performance}). The current-density
distribution was measured as a function of voltage and of axial
magnetic field. The results of the characterization were reported in
Refs.~\cite{Li:TM:2012, Moens:Thesis:2013}.

\subsection{Vacuum}

The Tevatron electron lenses were evacuated with 4~ion pumps (255~l/s
nominal total) and reached a typical residual pressure of
\q{10^{-9}}{mbar}. The insulating vacuum between the cold mass and the
warm beam pipe was \q{10^{-6}}{mbar}. Accessible components were baked
with heat tapes, whereas baking of inner surfaces was provided by
heating foils. In the LHC, the electron lens has to include, on each
side, a vacuum isolation module with gate valves, nonevaporable getter
(NEG) cartridges, pumps, and vacuum gauges. The length of each of
these modules is about 0.8~m. Surfaces need to be certified for
pressure and electron-cloud stability (electron-cloud multiplication
is suppressed when the solenoids are on).

\subsection{Electrical systems}

The TEL-2 gun and collector resistive solenoids required 340~A to
reach 0.4~T. The superconducting main solenoid yielded 6.5~T at
1780~A. The cathode, profiler electrode, anode bias, and collector
require 10-kV high-voltage power supplies.

A high-voltage modulator is used to pulse the anode and extract
current from the cathode. It needs to deliver 10~kV with a 10\%-90\%
rise time of 200~ns and a repetition rate of 35~kHz (3~times the
revolution frequency). This repetition rate would allow
synchronization of the electron beam with a subset of bunches for
tests and for direct comparison with the unaffected bunches. The
modulator requirements for collimation are less stringent than those
achieved with the TEL-2 stacked-transformer modulator for
bunch-by-bunch voltage adjustments in the
Tevatron~\cite{Pfeffer:JINST:2011}.

\subsection{Cryogenics}

Installation time is dominated by cryogenic integration, which would
be similar to that of a stand-alone magnet at 4.5~K. It requires at
least 3~months for warm-up, connection of the dedicated supply/return
interfaces with the distribution line (QRL), and cool-down. Electron
lenses may benefit from the dedicated rf refrigerator proposed for
installation in 2018. The Tevatron devices had static heat loads of
12~W for the helium vessel at 4~K and 25~W for the liquid nitrogen
shield. Nitrogen is not available in the LHC tunnel, but high-pressure
(20~bar) gaseous helium could be used instead for the shield. In the
Tevatron, the magnet string cooling system provided a flux of 90~l/s
of liquid helium. The quench protection system would have to be
integrated with that of the LHC.

\subsection{Diagnostics and controls}

The main superconducting solenoid incorporates 6 corrector magnets (1
long dipole positioned between 2 short dipoles in each plane) for the
alignment of the electron beam. Two stripline pickups (each one with
both horizontal and vertical plates) are positioned at the upstream
and downstream ends of the overlap region for accurate beam position
monitoring of both the long electron pulses and the short proton
pulses.  Sensitive loss monitors (such as scintillator paddles or
diamond detectors), positioned at the nearest aperture restrictions,
can be used to verify the relative beam alignment. In addition, if the
loss monitors are gated and synchronized with subsets of bunches, they
can provide a direct comparison between the intensity decay rates,
loss fluctuations, and halo diffusivities of bunches with and without
the electron-lens effect.

Monitoring of the electron beam profiles can be achieved with flying
wires or with fluorescent screens at low currents, and with pinhole
scans in the collector at high currents.  A direct measurement of the
halo population (through synchrotron light or induced fluorescence,
for instance), although not strictly necessary, would greatly benefit
this project and LHC operations in
general~\cite{Schmickler:CERN-Review:2012}. Biased electrodes on each
side of the overlap region can be used for clearing residual-gas ions
if necessary.

An electron lens test stand at CERN (possibly in collaboration with
the development of the ELENA electron cooler~\cite{Maury:IPAC:2013})
should be developed to characterize components and to develop
diagnostic techniques.

\subsection{Impedance of the electron-lens hardware}
\label{sec:imped}

Bunch structure and beam intensities in the LHC are very different
from those in the Tevatron. This translates into tighter requirements
on the electromagnetic impedance of the electron-lens hardware. (The
impedance effects of the electron beam itself are discussed in
Section~\ref{sec:e-beam-imped}.) In the Tevatron, the typical rms
bunch length was 2~ns and the bunch spacing was 395~ns. In the LHC,
the bunch spacing is 25~ns or 50~ns, and the typical bunch length is
0.3~ns.

The total longitudinal impedance of the Tevatron vacuum chamber and
components was a few ohms~\cite{Holmes:RunII:1998}, whereas the LHC
broad-band longitudinal impedance budget is only
\q{90}{m\ohm}~\cite{Bruening:LHC:2004}. TEL-2 stretched-wire
measurements showed several peaks between \q{0.1}{\ohm} and
\q{1}{\ohm} in the frequency range
0.1--1~GHz~\cite{Scarpine:AIP:2006}, confirming recent preliminary
simulations, which identified trapped modes in the electrode structure
(injection chamber, clearing electrodes, beam position monitors,
etc.)~\cite{Salvant:priv:2012}. The design of the electron-lens
electrodes will have to include provisions (such as rf shields) to
suppress wake fields, but this should not constitute a major
obstacle. The preliminary analysis of transverse impedances has not
raised any issues so far.

\section{Resources and schedule}

The construction cost of each of 2 electron lenses (one per beam) for
the LHC is estimated to be 2.5~M\$ in materials and 3.0~M\$ in
labor. This includes engineering, electron guns, resistive and
superconducting solenoids, vacuum chambers, electrodes, cabling,
instrumentation, and controls.

Construction of 2~devices would take about 3~years. Construction in
2015--2017 and installation during a long shutdown in 2018 is
technically feasible. Reuse of some of the Tevatron equipment, such as
superconducting coils, conventional solenoids, power supplies, and
electron guns, is also possible. Fermilab and BNL have the
capabilities and facilities for building the electron lens hardware.

Contributions in the areas of design, construction, commissioning,
numerical simulations, beam studies, and project management will be
specified in an agreement between CERN and US LARP.

\section{Alternative halo-removal schemes}

Hollow electron beam collimation is being evaluated in comparison with
other halo scraping techniques: tune modulation, damper excitation,
and beam-beam wire compensators.

Tune modulation with warm quadrupoles was used in HERA at DESY to
counteract the effects of power-supply
ripple~\cite{Bruening:EPAC:1994, Bruening:PRL:1996}. It was suggested
that this technique may allow one to excite a subset of particles in
tune space. Preliminary simulations with the \code{SixTrack} code
indicated that the halo cannot be removed as
selectively~\cite{Previtali:TM:2013}, but further investigations and
experimental tests are needed. Narrow-band excitations with the
transverse damper system were also proposed as a halo reduction
method~\cite{Hofle:priv:2012}. Beam tests may be possible in 2015
after resuming LHC operations. Both tune modulation and damper
excitation operate in tune space, where the core and the halo of the
beam are not necessarily separated.

Wire compensators for long-range beam-beam interactions are another
method one could use to manipulate the dynamic aperture in a
controlled way. It turns out that magnetically confined pulsed
electron beams may actually provide a better alternative not only for
scraping but also for long-range compensation, because they are not
electrically neutral (therefore requiring much less current), because
no material in close proximity with the circulating beam is involved,
and because their strength can be different for different
bunches~\cite{Valishev:TM:2013}.

\section{Conclusions}

Experimental and numerical studies were conducted to support the
conceptual design of a hollow electron beam collimator for the LHC, a
promising technique for controlled scraping of very intense
beams. This technique may be used in all cases in which material
damage, localized instantaneous energy deposition, or impedance limit
the use of conventional collimators.

The design was based on the experience of the existing Tevatron and
RHIC electron lenses. The expected halo cleaning performance and the
mitigation of undesired effects on the beam core were inferred from
the Tevatron experiments and from numerical tracking simulations. A
hollow electron gun with geometrical features and peak current yields
appropriate for the LHC was built and tested. To achieve a wide range
of halo removal rates, several electron beam pulsing modes were
studied. Hardware parameters and instrumentation options were
defined. No major obstacles were identified in the integration of the
devices in the LHC ring from the point of view of electromagnetic
impedance, mechanical engineering, or cryogenics. Required resources
were outlined. Studies of possible alternative schemes were
initiated. Further experimental tests may be possible with the RHIC
electron lenses to extend the Tevatron results. We also identified
other uses of electron lenses that could improve the performance of
the LHC: generation of tune spread for beam stabilization before
collisions; and long-range beam-beam compensation for luminosity
upgrade scenarios with small crossing angles.

Our studies suggest that hollow electron beam collimation could be
implemented in the LHC, if needed. This conceptual design report will
serve as the basis for a detailed technical design.

\begin{acknowledgments}
  This work has greatly benefited from the contributions and support
  of several people. In particular, the authors would like to thank
  O.~Aberle, A.~Bertarelli, F.~Bertinelli, E.~Bravin, O.~Br\"uning,
  G.~Bregliozzi, P.~Chiggiato, S.~Claudet, W.~Hofle, R.~Jones,
  Y.~Muttoni, L.~Rossi, B.~Salvant, H.~Schmickler, R.~Steinhagen,
  L.~Tavian, G.~Valentino (CERN), G.~Annala, G.~Apollinari, M.~Chung,
  T.~Johnson, I.~Morozov, E.~Prebys, G.~Saewert, V.~Shiltsev,
  D.~Still, L.~Vorobiev (Fermilab), R.~Assmann (DESY), V.~Kamerdzhiev
  (FZ J\"ulich), M.~Blaskiewicz, W.~Fischer, X~Gu (BNL), D.~Grote
  (LLNL), H.~J.~Lee (Pusan National U., Korea), S.~Li (Stanford U.),
  A.~Kabantsev (UC San Diego), T.~Markiewicz (SLAC), V.~Moens (EPFL),
  and D.~Shatilov (BINP).

 The authors would like to sincerely acknowledge the CERN colleagues who participated in the 2012 hollow e-lens review (G.~Arduini, O.~Bruening, S.~Claudet, K.~Cornelis, J.~Coupard, B.~Dehning, M.~Giovannozzi, B.~Goddard, E.~Jensen, W~.H\"{o}fle, A.~Grudiev, M.~Lamont, R.~Losito, R.~Schmidt, J.~Wenninger, M.~Zerlauth) and Steve Myers who chaired the panel that recommended the strategy adopted for the applications for the LHC.

\end{acknowledgments}


\clearpage

\printtables

\clearpage

\printfigures

\clearpage


\end{document}